# Kolmogorov-Arnold Neural Networks for High-Entropy Alloys Design

Yagnik Bandyopadhyay,[1] Harshil Avlani,[1,2] and Houlong L. Zhuang[1,*]

[1]School for Engineering of Matter Transport and Energy, Arizona State University, Tempe, AZ 85287, USA

[2]Basis Chandler High School, Chandler, AZ 85286, USA

*zhuanghl@asu.edu

**Abstract**

A wide range of deep learning-based machine learning techniques are extensively applied to the design of high-entropy alloys (HEAs), yielding numerous valuable insights. Kolmogorov-Arnold Networks (KAN) is a recently developed architecture that aims to improve both the accuracy and interpretability of input features. In this work, we explore three different datasets for HEA design and demonstrate the application of KAN for both classification and regression models. In the first example, we use a KAN classification model to predict the probability of single-phase formation in high-entropy carbide ceramics based on various properties such as mixing enthalpy and valence electron concentration. In the second example, we employ a KAN regression model to predict the yield strength and ultimate tensile strength of HEAs based on their chemical composition and process conditions including annealing time, cold rolling percentage, and homogenization temperature. The third example involves a KAN classification model to determine whether a certain composition is an HEA or non-HEA, followed by a KAN regressor model to predict the bulk modulus of the identified HEA, aiming to identify HEAs with high bulk modulus. In all three examples, KAN either outperform or match the performance in terms of accuracy such as $F_1$ score for classification and Mean Square Error (MSE), and coefficient of determination ($R^2$) for regression of the multilayer perceptron (MLP) by demonstrating the efficacy of KAN in handling both classification and regression tasks. We provide a promising direction for future research to explore advanced machine learning techniques, which lead to more accurate predictions and better interpretability of complex materials, ultimately accelerating the discovery and optimization of HEAs with desirable properties.

**Introduction**

High entropy alloys (HEAs) are a special class of materials characterized by their unique composition of multiple principal elements[1, 2]. Unlike traditional alloys dominated by one or two elements, HEAs benefit from the high entropy of mixing, leading to numerous desirable mechanical properties such as hardness, strength, ductility, and bulk modulus [3, 4]. Because HEAs comprise multiple principal elements, the number of possible compositions is much higher than that of traditional alloys. This complexity poses significant challenges in the design and optimization of HEA compositions, necessitating advanced computational methods to explore the vast compositional space effectively. Given the vast compositional landscape and the need to predict material properties accurately, machine learning (ML) has emerged as a critical tool in the design of HEAs [5-7]. By learning from existing data, ML models can predict the properties of new HEA compositions efficiently and reduce the time and cost associated with experimental trials. This approach allows for the rapid screening of potential alloys and accelerates the discovery of new materials [8-11] with desired properties. Multi-layer perceptron (MLP) (also known as deep neural networks) have been widely used to explore non-linear relationships between compositional features and material properties, leading to the design of optimized HEA compositions and facilitating material discovery [12, 13]. For example, an MLP regression is used to predict the phase constitution in HEA systems [14, 15]. Another study uses a data-driven approach that combines ML, experimental design, and feedback from experiments to accelerate the search for multi-component alloys with specific target properties [16]. Despite their advantages, MLPs have a few challenges when applied to the design of HEAs. One of the challenges is model confidence i.e., understanding when the model gives accurate predictions and when it may not. Also, the black-box nature of MLPs often hinders the interpretability of the model, making it challenging to derive meaningful insights about the underlying material science [17].

To address some of the limitations of traditional neural networks like MLPs, Kolmogorov-Arnold Neural Networks (KAN) have been proposed recently [18]. KAN leverages the Kolmogorov-Arnold representation theorem [19-21], stating that if a function is multivariate continuous function on a bounded domain, then the function can be written as a finite composition of continuous functions of a single variable and the binary operation of addition. Despite their elegant mathematical interpretation, KAN is essentially combinations of splines [22-24] and MLPs, leveraging their respective strengths and avoiding their respective weaknesses. To learn a function accurately, a model should not only learn the compositional structure (external degrees of freedom) but also approximate well the univariate functions (internal degrees of freedom). KAN structured MLPs on the outer layers and splines on the inner layers. As a result, KAN can learn features (thanks to their external similarity to MLPs) and optimize these learned features to great accuracy (thanks to their internal similarity to splines). This unique architecture allows KAN to model complex functions more efficiently and with greater interpretability than conventional neural networks.

The original KAN paper demonstrates various toy examples to illustrate the power of KAN in terms of accuracy and interpretability and their advantage over MLPs. KAN is also used in a physics problem related to Anderson Localization [25], where KAN is applied to numerical data generated from quasiperiodic tight-binding models to extract their mobility edges. There have also been recent applications of KAN in solving mechanics problems [26], time series forecasting, and leveraging their adaptive activation functions for enhanced predictive modeling in real-world satellite traffic forecasting tasks [27].

Our work is a prospective study that applies KAN to the design and optimization of HEAs. We employ KAN on three different datasets as examples previously used to classify HEAs from non-HEAs or to predict mechanical properties of HEAs [3, 28] such as universal tensile strength (UTS), yield strength (YS), and bulk modulus and compare the model performance primarily with MLPs.

**Methods**

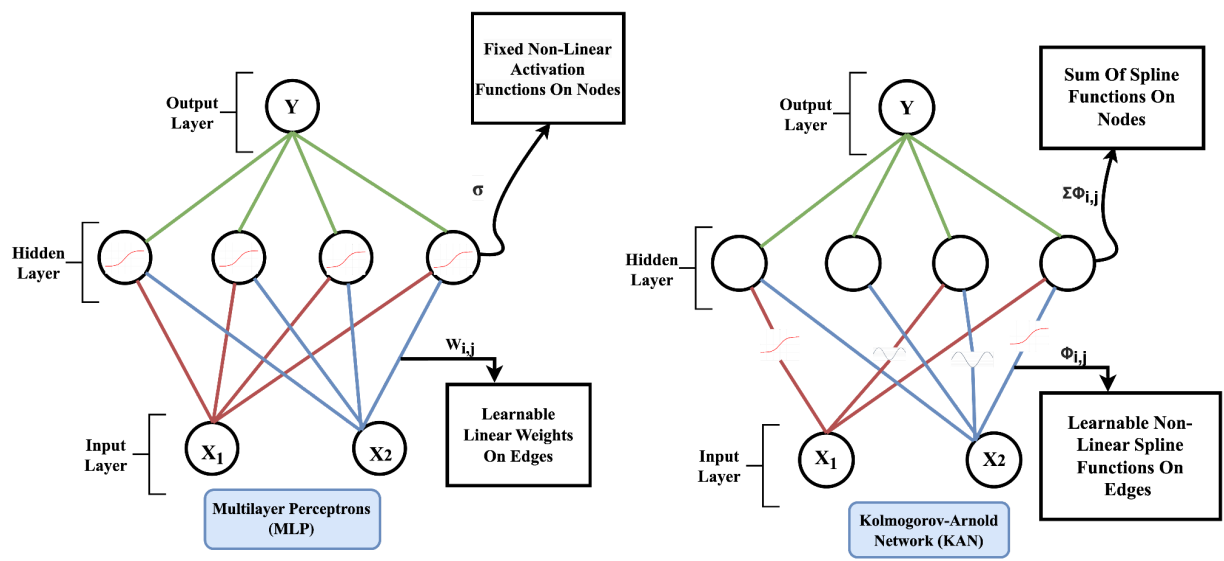

**Figure 1**. Illustrations of (a) MLP and (b) KAN architectures. The illustration of KAN is adopted from Ref. [18].

Figure 1 illustrates the architectures of MLP and KAN. In MLP, the trainable parameters are the weights and biases of a network. Each node in a layer is connected to nodes in the subsequent layer. The weights and biases determine the significance of specific connections during the model's learning process. Activation functions, such as ReLU and sigmoid, introduce non-linearity into the model's regression, enhancing its ability to learn complex patterns. Like MLP, KAN has fully connected structures. However, while MLP places fixed activation functions on *nodes* ("neurons"), KAN places learnable activation functions on *edges* ("weights"). As a result, KAN has no linear weight matrices at all: instead, each weight parameter is replaced by a learnable 1D function parameterized as a spline. KAN's nodes simply sum incoming signals without applying any non-linearities. Splines introduce non-linearity into the data by using piecewise polynomial functions that can model high-dimensional data as polynomial 1D functions. This allows KAN to capture more intricate data relationships compared to MLP. Splines also offer transitions between different resolutions, allowing for better and more accurate learning in regions that require more detail. However, splines also have the problem of the curse of dimensionality, meaning they cannot handle high-dimensional data effectively. KAN integrates splines within an MLP framework in a way that provides the advantage of learning both the compositional structure (external degrees of freedom), which approximates the high-dimensional aspects of the data, and the univariate functions (internal degrees of freedom) with the help of splines [18].

In both MLP and KAN, there are a number of parameters that can be tuned to optimize the performance of the corresponding neural networks. For example, MLP consists of setting different numbers of layers and nodes in each layer. MLP and KAN differ significantly in their architectural and operational features. MLP typically consists of multiple layers, each with a fixed number of nodes, where the optimization algorithm is often based on gradient, and the activation functions are fixed, such as ReLU or sigmoid. MLP is trained using a specified batch size and a set number of epochs to minimize the loss function. In contrast, KAN also has multiple layers, but the number of nodes at each layer can vary, and they employ learnable activation functions along the edges rather than fixed linear weights. KAN uses a learning rate and regularization parameter to control the training process, and they adaptively adjust the grid to fine-tune the model. Additionally, KAN is trained over a series of steps rather than epochs, allowing for more dynamic adjustments during the training process. For validation, we employ shuffle split and k-fold cross-validation (CV) techniques. In KAN, the activation function $\varphi(x)$ is the sum of the basis function $b(x)$ and the spline function: $\varphi(x) = w_b \cdot b(x) + w_s \cdot \text{spline}(x)$, where $w_b$ and $w_s$ are weight factors controlling the magnitude of the $b(x)$ and spline(x) functions. $b(x) = x/(1 + e^{-x})$ (also known as silu($x$)) and spline($x$) is parametrized as a linear combination of B-splines such that $\text{spline}(x) = \sum_{i=0}^{G+k-1} c_i B_i(x)$ where, G represents the grid size, k is the degree of the spline function, and $c_i$ are trainable parameters. In principle, $w_b$ and $w_s$ are redundant since they can be included into the $b(x)$ and spline(x) functions [18]. However, we still account for these factors (which by default are trainable) to better control the overall magnitude of the activation function. In both MLP and KAN, we used a random seed of zero to split the training and test data.

Apart from the common ways like feature importance, KAN provides intuitive visualization to understand the learning ability of the model by visualizing the spline functions. KAN offers interpretability that MLPs cannot provide with the help of features which can plot the learnable spline function at each of the edges of the KAN model or use symbolic regression which provides symbolic formulas of how the input features relate to the target variable/output. In this work, we implement the MLP and KAN architectures built on the datasets used in Refs. [29-31].

**Results and Discussion**

To demonstrate the applications of KAN for HEAs, we examine three recent example studies [29-31] where MLPs are applied to design HEA. The machine learning tasks in the three studies fall into the categories of classification, regression, or a combination of both. Our objectives here are twofold: (i) We compare our KAN method against other techniques, especially the MLP model reported in the literature. (ii) We aim to assess how interpretable the outcomes from our KAN approach are compared to these recently published works. By analyzing these case studies, we aim to showcase the advantages and potential of leveraging KAN for designing HEAs, particularly in terms of computational performance and result interpretability.

**Example One: Classification Model**

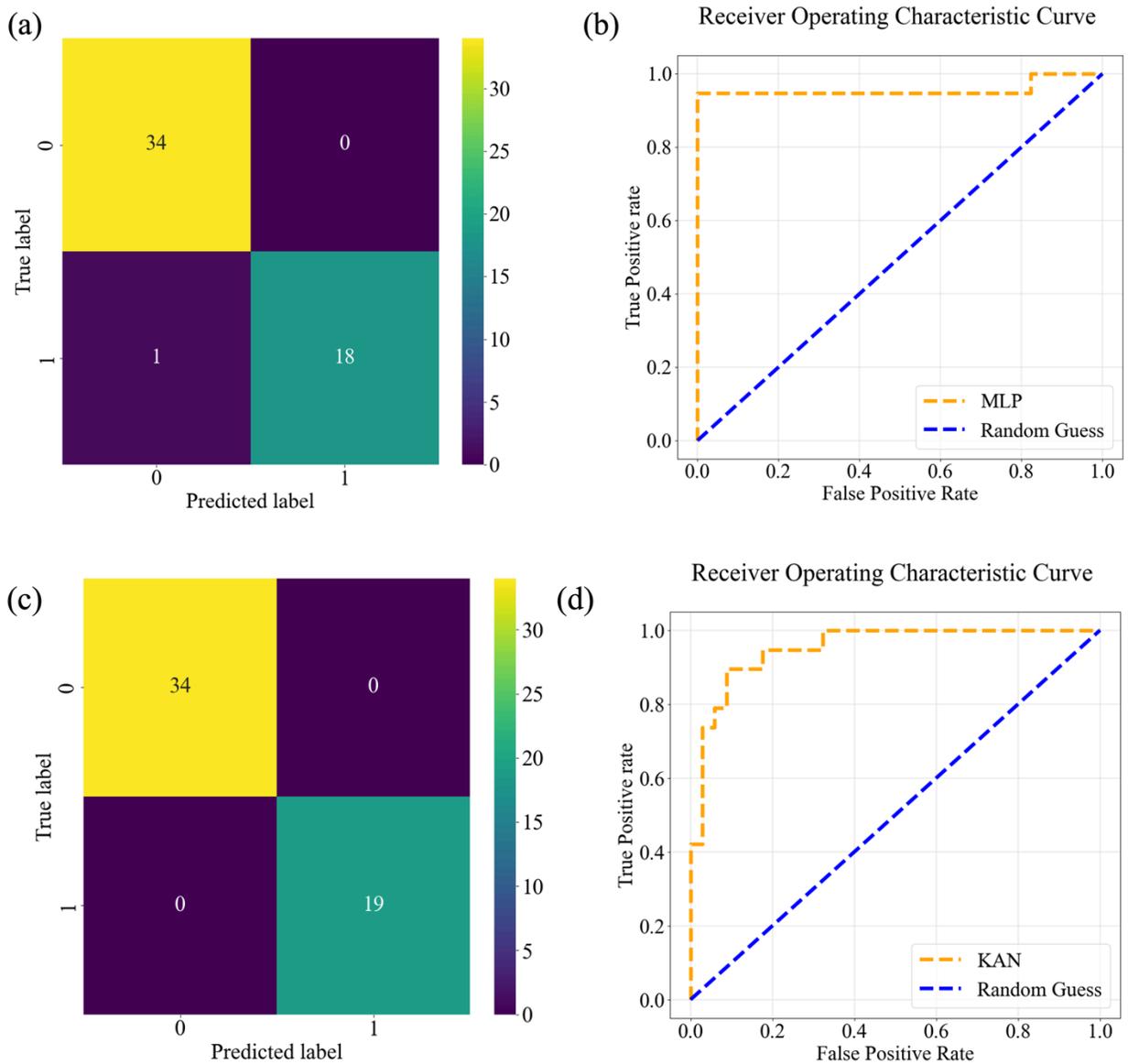

**Figure 2.** (a) Confusion Matrix of example one using MLP. (b) Receiver Operating Characteristic Curve of example one using MLP. (c) Confusion Matrix of example one using KAN. (d) Receiver Operating Characteristic Curve of example one using KAN.

The primary motivation of this example study [31] is to predict the phases of novel HEAs using machine learning. The input features for the machine learning calculations include elemental attributes such as valence electron concentration of HECCs. Two types of machine learning models: MLPs and support vector machines (SVM) are used. The models are trained using data from density functional theory calculations and also from experiment. The final accuracy obtained with MLP and SVM models are 0.982 and 0.944, respectively. The trained MLP and SVM models are then used to predict the probabilities of single phase for 90 novel HECC samples. The classification threshold is set to 0.5 and 38 out of 90 samples are predicted as single-phase HECCs. These samples have not been reported in experiments before. Thirteen HECC candidates with high and low single-phase probabilities are then selected to validate the ML models through experiments. In case of discrepancies between the MLP and SVM model, the prediction of the MLP model is selected, as it is more accurate. Nine candidates, such as $(HfMoNbTaV)C_5$ are predicted to exhibit the single-phase face-centered cubic (FCC) structure, while the other four candidates (e.g., $(CrHfTaVW)C_5$) are predicted to exhibit multiple-phase microstructures. These predictions from ML models align well with the experimental results obtained in the same study.

In this work, we first train our own MLP model on the same HECC dataset to obtain a better comparison with the MLP model in Ref. [31]. We then deploy the KAN model, which has a single hidden layer with three nodes, to the dataset to evaluate its accuracy and F1 score in predicting the phases of HECC candidates. Here, the F1 score is a metric used to evaluate the accuracy of a model by calculating the harmonic mean of precision and recall. Precision is measured as the ratio of true positive (TP) to the sum of true positive and false positive predictions (TP + FP). Recall is the ratio of true positive predictions to the sum of true positive and false negative predictions (TP+FN). Figure 2 shows the confusion matrix, and the receiver operating characteristic (ROC) curve obtained from training the MLP and KAN models. The results and comparisons between MLP (Ref. [31]), MLP (this work), and KAN are summarized in Table 1. Our MLP model, despite having slightly different configurations in the neurons within the hidden layers, achieves the same accuracy, F1 score, and the area under the curve (AUC) as those reported in Ref. [31]). Additionally, the single compound (HfTiTaZr)$C_5$ whose phase our MLP model fails to predict accurately is the same compound that is not predicted correctly in the reference. By contrast, the confusion matrix resulted from the KAN model (Fig. 1c) shows the absence of any FPs or FNs during prediction, and the AUC of the ROC curve is 1 (Fig. 1d). In other words, we achieve a maximum accuracy of 1.0, which surpasses the best-optimized MLP model's accuracy of 0.982. We also perform cross-validation (CV) calculations with 25 shuffle splits and obtain a mean F1 score of 0.83. Additionally, the score of our KAN model is also higher than that of the MLP counterpart.

We apply the trained model to the 90 unknown HECC materials to predict their phases. Our KAN classification model predicted 45 out of the 90 materials are single-phase, while the MLP model from the Ref. [31] predicted only 38 single-phase HECCs. However, upon comparing the 13 experimentally determined phases of compounds from [31] with those predicted by the KAN model, we also identified another study [32] that experimentally determine the phases of 22 out of the 90 compounds, four ((CrHfNbTaTi)$C_5$, (CrHfTaTiZr)$C_5$, (CrHfNbTaV)$C_5$, (MoNbTaVZr)$C_5$)) of which overlap with those in Ref. [31]. Table 2 summarizes the model predictions of both KAN and MLP and compares them to the experimental results of [31]and [32]. KAN's predictions match 8 out of the 13 compounds listed in Ref. [31] and 8 out of the 22 in Ref. [32]. However, among the four compounds common to both studies, two show conflicting results. Specifically, (CrHfNbTaV)$C_5$ and (MoNbTaVZr)$C_5$ are classified as single-phase compounds in Ref. [31], while Ref. [32] reports them as multi-phase compounds. This inconsistency highlights the challenges in evaluating the accuracy of the KAN model for new HECCs and underscores the need for further investigation. Nevertheless, there is a significant challenge and a scope for future work in terms of avoiding overfitting and providing better generalizability of the model.

Table 1. Comparison of MLP and KAN performance in a classification task of HEAs.

| Model | Layer Width | Number of parameters | Accuracy | F1 Score | AUC |
|---|---|---|---|---|---|
| MLP* | (12, 6, 6, 2) | 134 | 0.98 | 0.97 | 0.96 |
| MLP+ | (12, 20, 20, 2) | 722 | 0.98 | 0.97 | 0.96 |
| KAN+ | (12, 3, 2) | 761 | 1.0 | 1.0 | 1.0 |

*Ref. [31]
+ This work

Table 2. MLP and KAN models applied to 31 HECC compounds and compared with available experimental results. The compounds whose prediction could be compared with both experimental results in Refs [31, 32] are highlighted in boldface.

| Compounds | MLP | KAN | Experimental Result Ref. [32] | Experimental Result Ref. [35] |
|---|---|---|---|---|
| (CrHfNbTaZr)C$_5$ | Single | Multiple | | Single |
| (CrMoNbTaZr)C$_5$ | Multiple | Single | | Multiple |
| **(CrHfNbTaTi)C$_5$** | **Single** | **Multiple** | **Single** | **Single** |
| (CrMoNbTaTi)C$_5$ | Single | Single | | Multiple |
| **(CrHfTaTiZr)C$_5$** | **Single** | **Multiple** | **Single** | **Single** |
| (CrMoTaTiZr)C$_5$ | Multiple | Multiple | | Single |
| (CrHfNbTiZr)C$_5$ | Single | Multiple | | Single |
| (CrHfMoVW)C$_5$ | Multiple | Single | Multiple | |

| Composition | | | | |
|---|---|---|---|---|
| (CrNbTaTiZr)C$_5$ | Single | Single | | Single |
| (MoNbTaTiW)C$_5$ | Single | Single | Single | |
| (CrHfTaTiV)C$_5$ | Single | Multiple | | Multiple |
| (CrHfNbTiV)C$_5$ | Single | Single | | Multiple |
| (CrMoNbTiV)C$_5$ | Multiple | Single | | Multiple |
| (CrNbTaTiV)C$_5$ | Single | Single | | Single |
| (CrHfTiVZr)C$_5$ | Multiple | Multiple | | Multiple |
| (CrMoTiVZr)C$_5$ | Multiple | Single | Multiple | |
| (CrTaTiVZr)C$_5$ | Single | Single | | Multiple |
| (MoTiVWZr)C$_5$ | Multiple | Multiple | Multiple | |
| (CrMoTaTiV)C$_5$ | Single | Multiple | | Single |
| (CrHfTaVW)C$_5$ | Multiple | Multiple | Multiple | |
| **(CrHfNbTaV)C$_5$** | **Single** | **Multiple** | **Single** | **Multiple** |
| (CrMoNbTaV)C$_5$ | Single | Single | | Single |
| (CrHfTaVZr)C$_5$ | Single | Multiple | | Multiple |
| (CrHfNbVZr)C$_5$ | Multiple | Multiple | | Multiple |
| (CrMoNbTiZr)C$_5$ | Multiple | Single | | Multiple |
| (CrNbTaVZr)C$_5$ | Single | Single | | Multiple |
| (HfNbTaVW)C$_5$ | Single | Single | Single | |
| (HfMoNbTaV)C$_5$ | Single | Single | Single | |
| **(MoNbTaVZr)C$_5$** | **Single** | **Single** | **Single** | **Multiple** |
| (NbTaVWZr)C$_5$ | Single | Single | Single | |

| (HfMoNbTaZr)C$_5$ | Single | Single | Single | |

**Example Two: Regression Model**

The primary objective of Ref. [30] is to develop a regression model capable of accurately predicting mechanical properties like and ultimate tensile strength (UTS) and yield strength (YS) of HEAs. The HEA dataset is composed of the chemical compositions and process conditions that include the homogenization temperature, annealing temperature, annealing time, and cold rolling percentage of 501 HEAs, aggregated from experimental studies. This dataset is used to train an MLP, which achieves a test root mean square error (RMSE) of 157 and a R² of 0.828 for YS prediction, and a test RMSE of 146 and a R² of 0.763 for UTS prediction, averaged over 100 holdout cross-validations (CVs).

We apply the KAN model to the same HEA dataset to compare its performance and size with the MLP model. Since comprehensive performance metrics for the MLP model are not provided in Ref. [30], we perform our own set of MLP model calculations using the same HEA dataset.

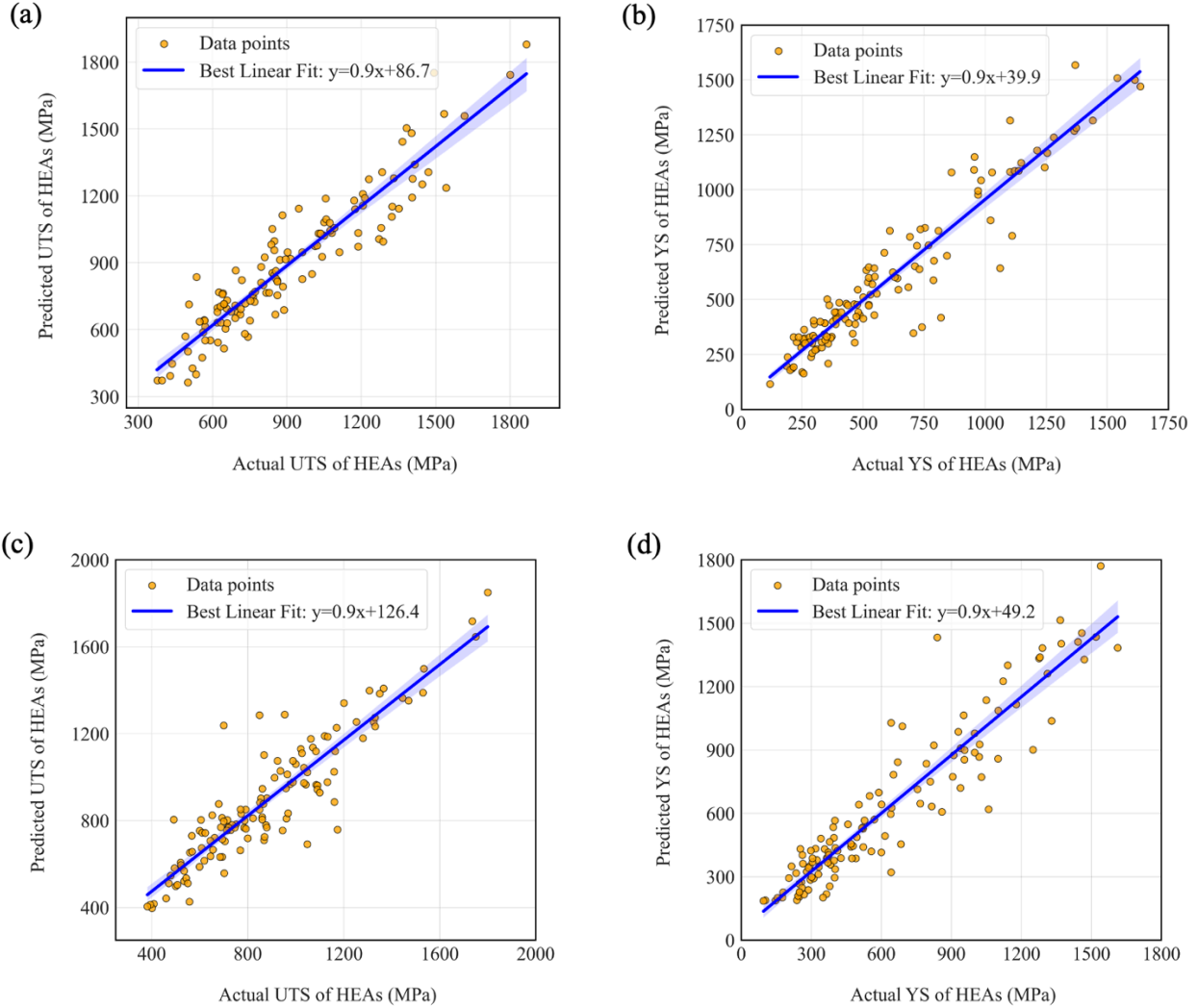

**Figure 3.** (a) Actual and predicted Ultimate Tensile Strength using MLP, (b) Actual v/s Predicted yield strength using MLP, (c) Actual and predicted ultimate tensile strength using KAN, and (d) Actual and predicted yield strength using KAN.

The purpose of one of these models is to establish the best performance for MLP, and the other purpose is to establish MLP performance on a model with similar parameter counts to KAN. We first standardize our data, and then refine the KAN and MLP models through a combination of manual tuning and the Optuna [33] hyperparameter tuning framework. Figures 3 and 4 show the actual UTS and YS of the HEAs compared to the predicted values using KAN and MLP, respectively. Our MLP model outperforms the previous work, obtaining a CV test RMSE of 138 and CV test R² of 0.780 in UTS prediction, and a CV test RMSE of 140 and CV test R² of 0.847 in YS prediction. In YS prediction, MLP greatly outperforms KAN, which achieves a CV test RMSE of 155 and CV test R² of 0.810. However, in UTS prediction, MLP only slightly outperforms KAN, achieving a CV test RMSE of 142 and a CV test R² of 0.771. This trend continues when considering best model performance, with MLP achieving a test RMSE of 113 and a test R² of 0.873 and KAN achieving a test RMSE of 126 and test R² of 0.822 in UTS prediction. In the YS prediction, MLP achieves a test RMSE of 111 and test R² of 0.900 and KAN achieves a test RMSE of 136 and test R² of 0.873.

While MLP achieves higher accuracies overall, particularly in YS prediction, compared to KAN, it requires significantly more parameters and longer training times, as shown in Tables 3 and 4. When MLP models with comparable parameter counts are tested, much lower accuracy is achieved compared to KAN, achieving an RMSE of 162 and R² of 7.06 in UTS prediction and a RMSE of 189 and R² of 0.798.

**Table 3.** Comparison of MLP and KAN performance in UTS Regression

| Model | Parameters | Test RMSE | Train RMSE | Test R² | Train R² | CV Test RMSE | CV Test R² |
|---|---|---|---|---|---|---|---|
| MLP* | -- | -- | -- | -- | -- | 146 | 0.763 |
| MLP+ | 641 | 130 | 130 | 0.827 | 0.814 | 162 | 0.706 |
| MLP+ | 26323 | 113 | 52 | 0.873 | 0.970 | 138 | 0.780 |
| KAN | 634 | 126 | 122 | 0.822 | 0.840 | 142 | 0.771 |

When MLP models with comparable parameter counts are tested they achieve much lower performances compared to KAN, achieving an RMSE of 162 and R² of 7.06 in UTS prediction and a RMSE of 189 and R² of 0.798.

**Table 4.** Comparison of MLP and KAN performance in YS regression.

| Model | Parameters | Test RMSE | Train RMSE | Test $R^2$ | Train $R^2$ | CV Test RMSE | CV Test $R^2$ |
|---|---|---|---|---|---|---|---|
| MLP* | -- | -- | -- | -- | -- | 157 | 0.828 |
| MLP+ | 1089 | 155 | 152 | 0.798 | 0.833 | 189 | 0.798 |
| MLP+ | 32885 | 111 | 66 | 0.900 | 0.969 | 140 | 0.847 |
| KAN | 1085 | 136 | 122 | 0.873 | 0.886 | 155 | 0.810 |

* [30]
+ This work

**Example Three: Classification and Regression Models**

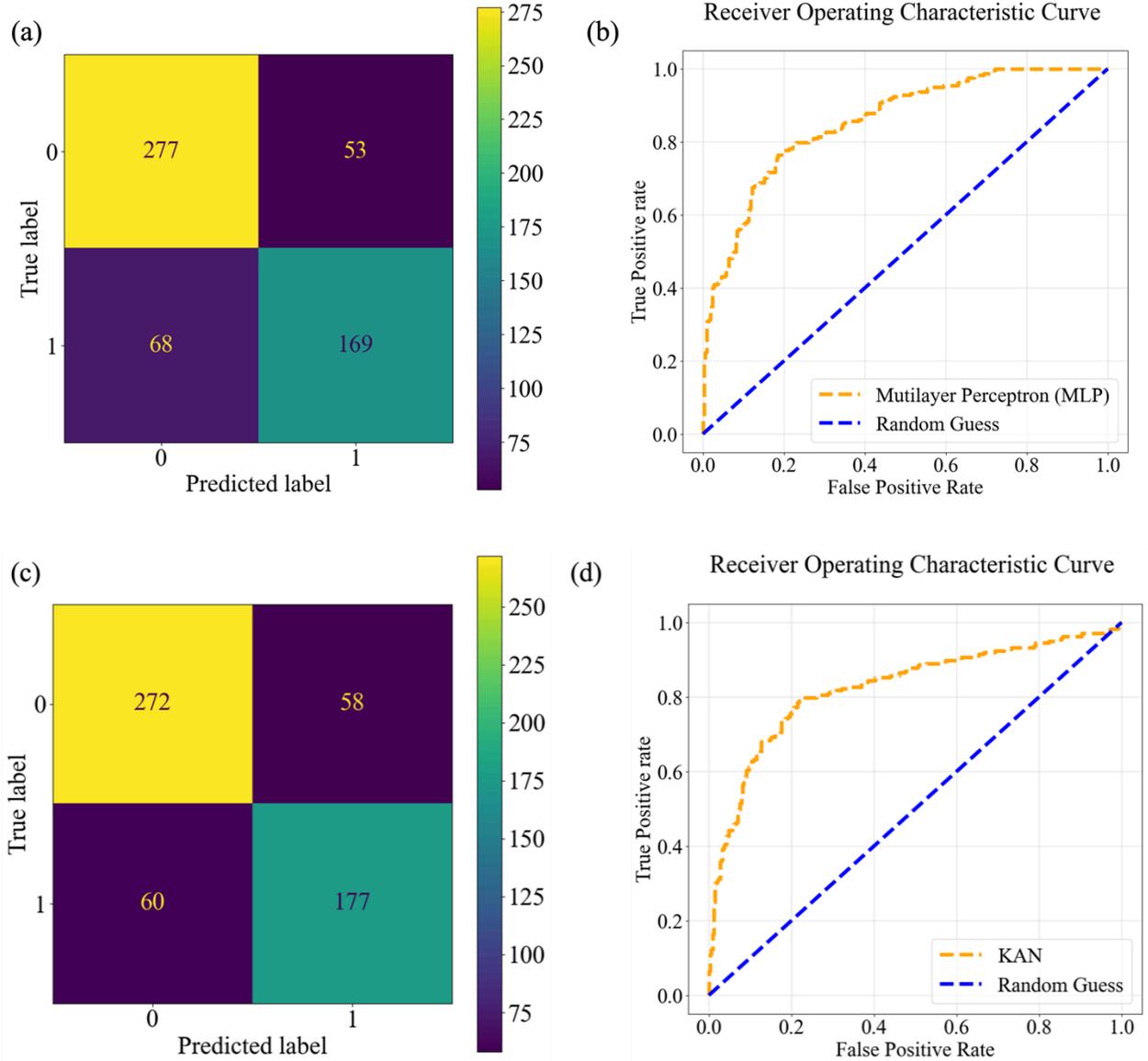

**Figure 4**. (a) Confusion matrix for the classification of HEA and Non-HEA using MLP, (b) Receiver Operating Characteristic Curve for classification of HEA and Non-HEA using MLP, (c) Confusion matrix for the classification of HEA and Non-HEA using KAN, (d) Receiver Operating Characteristic Curve for classification of HEA and Non-HEA using KAN.

The motivation in the work of Ref. [29] is to design HEAs with high bulk modulus using machine learning. Bulk modulus offers a quantitative measure of compressibility of an alloy. Materials with a high bulk modulus are well-suited for applications requiring operation under substantial pressure from all directions, such as machines functioning in deep ocean environments like submarines, pipelines, and torpedoes [29, 34]. Developing HEAs with a high bulk modulus therefore opens new opportunities in the field of HEAs. It is widely recognized that a high bulk modulus is associated with high strength [29, 35, 36]. The bulk modulus k is defined as $k = -v\frac{dP}{dv}$, where v is initial volume of the material, and dP and dv represents the change in pressure and volume respectively. The ML work in Ref. [29] is performed in several steps. First, the constituent elements with their mole fractions are extracted from the raw data and labeled the HEAs as 1 and non-HEAs as 0. The phase data of the compounds was converted into binary vectors for input. Next, each chemical composition is classified as either HEA or non-HEA using a machine learning model for classification. Once classification is complete, the bulk modulus of HEA compounds are predicted using a regression model. The final stage involves identifying the HEA composition with the highest or desired bulk modulus that is user dependent. The study leads to a predictive model which determines the optimum elemental composition of HEAs for maximizing bulk modulus. An HEA classification is created to predict if a certain elemental composition is an HEA, and also their bulk modulus using a regression model. First, the study applies various classification models like Gradient boosting [37] classification, support vector classification [38], etc., and then various regression models like Random forest regression [39], Lasso regression [40], Linear regression [41], etc. to predict the Bulk Modulus of the HEA. Finally, the performance of the models are compared to check which one performs the best. The two best models obtained are gradient boosting [37] for the classification model, with a classification accuracy of 0.77, and the LASSO regression [40] for bulk modulus prediction, which achieves an $R^2$ value of 0.99. Here $R^2$ is calculated as $R^2$ = 1-(SSE/SST), where SSE refers to the sum of squared errors that is the sum of the squared differences between the actual values and the predicted values from the regression model. SST is the total variation in the dependent variable. It is calculated by summing the squared differences between each actual value and the mean of all the values.

Ref. [29] does not use any MLP-based model to train either classification or regression task. To fairly compare the performance of KAN in both classification and regression, this study also develops MLP classification and regression models. These models are optimized for the best accuracy and then compared against the KAN model. The optimized KAN model for classification had three hidden layers with 20, 10, and 6 neurons, respectively, and for the regression, only one hidden layer is used with a width of 15 neurons. For the MLP, the optimized classification model had three hidden layers with widths of 104, 24, and 16, respectively, and the regression model also had three layers with widths of 175, 84, and 43 neurons, respectively. We use the Optuna [33] package along with manual tests to obtain the best performing model and optimized hyperparameters for both the KAN and MLP classification and regression models. Table 5 compares the performance of the KAN classification model against the MLP classification model and the best performing classification model used in Ref. [29]. Using the MLP model for classification, we obtain an optimum accuracy of 0.77, matching previous works. The KAN model, however, achieves the best test accuracy of 0.79 and an F1 score of 0.75. Figure 4 shows the confusion matrix and the ROC curve to classify HEAs and non-HEAs using MLP and KAN classification models, respectively. As can be seen, the KAN classification once again slightly outperforms the MLP classification model.

For the regression model, the MLP model obtained an MSE value of 13.85 and an $R^2$ value of 0.98, matching the accuracy of the previous work. The KAN model also achieves an $R^2$ value of 0.98. Table 6 compares the model performance of the KAN regression, MLP regression, and the best performing regression model from [29]. Additionally, a shuffle split and K-Fold cross-validation with 25 splits were performed. The shuffle split yielded an $R^2$ value of 0.97 for the KAN model, and the K-Fold validation yielded an $R^2$ value of 0.98 for the KAN model. For the MLP, the K-Fold $R^2$ was 0.98, matching the KAN performance. KAN therefore shows better prediction accuracy for the classification model compared to both MLP and the gradient boosting classification model that achieves a maximum accuracy of 0.77. For the regression model, KAN performs at par with the MLP regression model. Figure 5 compares the actual and predicted Bulk modulus values of HEA using MLP and KAN, respectively. Both MLP and KAN fit the line really well, and are almost similar in terms of prediction accuracy of Bulk Modulus.

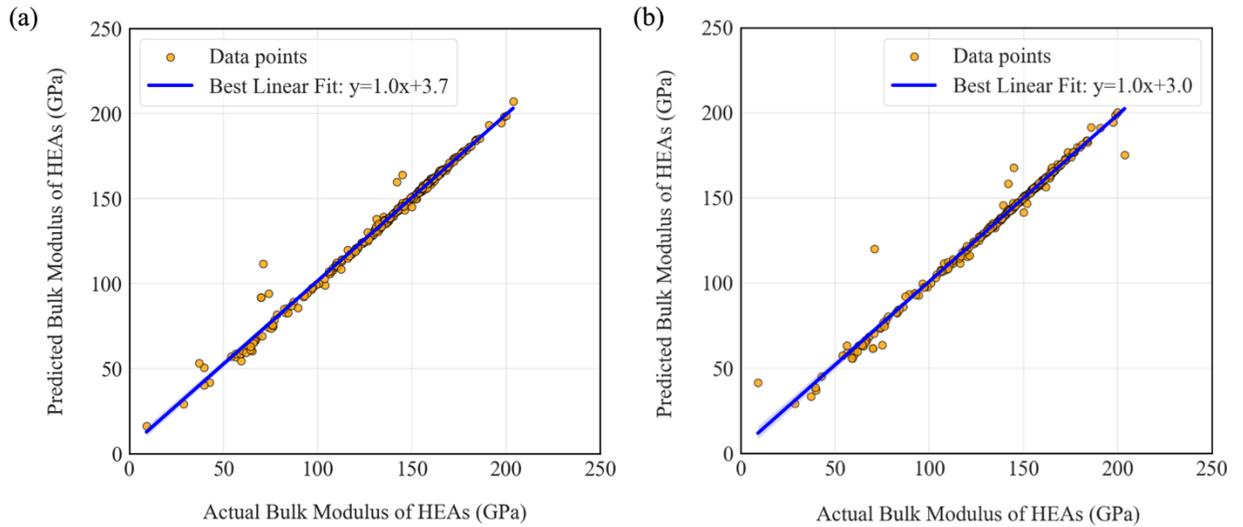

**Figure 5.** (a) Actual and predicted bulk modulus of HEAs using MLP, (b) Actual and predicted bulk modulus of HEAs using KAN.

**Table 5.** Comparison of MLP and KAN performance in a combination of classification of HEAs.

| Model | Layer Width | # of parameters | Train Accuracy | Test Accuracy | Train F1 score | Test F1 Score | Train AUC | Test AUC |
|---|---|---|---|---|---|---|---|---|
| GBC* | - | - | 0.99 | 0.77 | 0.99 | 0.72 | 0.99 | 0.76 |
| MLP+ | (79,104,24,16,2) | 11,274 | 0.90 | 0.78 | 0.88 | 0.73 | 0.90 | 0.77 |
| KAN+ | (79,20,10,6,2) | 33,374 | 0.95 | 0.79 | 0.94 | 0.75 | 0.95 | 0.79 |

**Table 6.** Comparison of MLP and KAN performance in a combination of regression tasks of HEAs.

| Model | Layer Width | # of parameters | Train MSE | Test MSE | Train $R^2$ | Test $R^2$ | Shuffle Split -25 Test $R^2$ | K-Fold Split -25 Test $R^2$ |
|---|---|---|---|---|---|---|---|---|
| Lasso Regression* | - | - | 1.68 | 26.98 | 0.99 | 0.98 | - | - |
| MLP[+] | (81, 175, 84, 43,1) | 32,833 | 2.67 | 15.91 | 0.99 | 0.98 | 0.97 | 0.99 |
| KAN[+] | (81, 15, 1) | 22,156 | 0.23 | 20.69 | 0.99 | 0.98 | 0.97 | 0.98 |

* [29]
[+] This work

**Interpretability of KAN Spline vs MLP activation Curve**

To evaluate the interpretability of KAN relative to a standard MLP, we analyze three different examples: a classification task, a regression task, and a combined classification-regression task. In each case, we select a few representative input feature nodes and compare the output functions generated at each connection in the KAN architecture with their MLP counterparts. For the MLP, we train models with equivalent complexity (same number of hidden layers and nodes) for an easier intuitive comparison. To visualize the outputs, we compute the MLP output at each connection using the weight (W) and bias (b) parameters and apply appropriate activation functions (i.e., softmax [42] or rectified linear unit (ReLU) [22, 43]) at each node connection. The general form of the equations is as follows:

$$Y_{i,j} = X_i * W_{i,j} + b_i$$

$$Z_{i,j} = \sigma(Y_{i,j}),$$

where σ is an activation function (Softmax for classification and ReLU for regression). We then select a connection from each model to plot the MLP outputs and compare these plots to the KAN spline function plots for the same connections.

**Example 1: Classification**

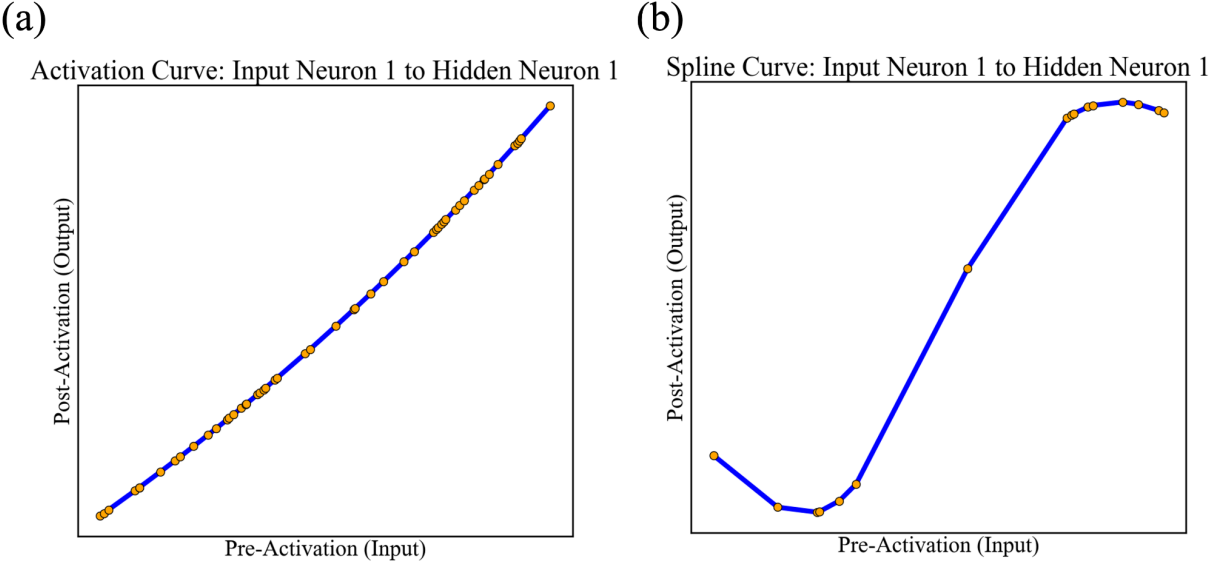

**Figure 6.** (a) MLP activation curve, (b) KAN spline curve of the Classification of HECC model for Input node 1 to output node 1 connection.

In this case, we visualize the output for both MLP and KAN models at certain node connections chosen randomly. Figure 6 reveals that while the MLP captures the overall trend, i.e., both the plots have a similar linear line plot before the observable curves at the beginning and end of the plot for KAN. KAN introduces additional complexity, particularly in the non-linear regions. For instance, the MLP output (Fig. 6a) shows a smooth linear behavior, while the plot (Fig. 6b) from the KAN model exhibits nonlinearity (like sine curve) due to its trainable spline functions, which can help the model learn complex relationships among the features. This enhanced nonlinearity allows KAN to make more accurate classifications as seen in this example where KAN instead of MLP can get rid of the misclassified data point.

**Example 2: Regression**

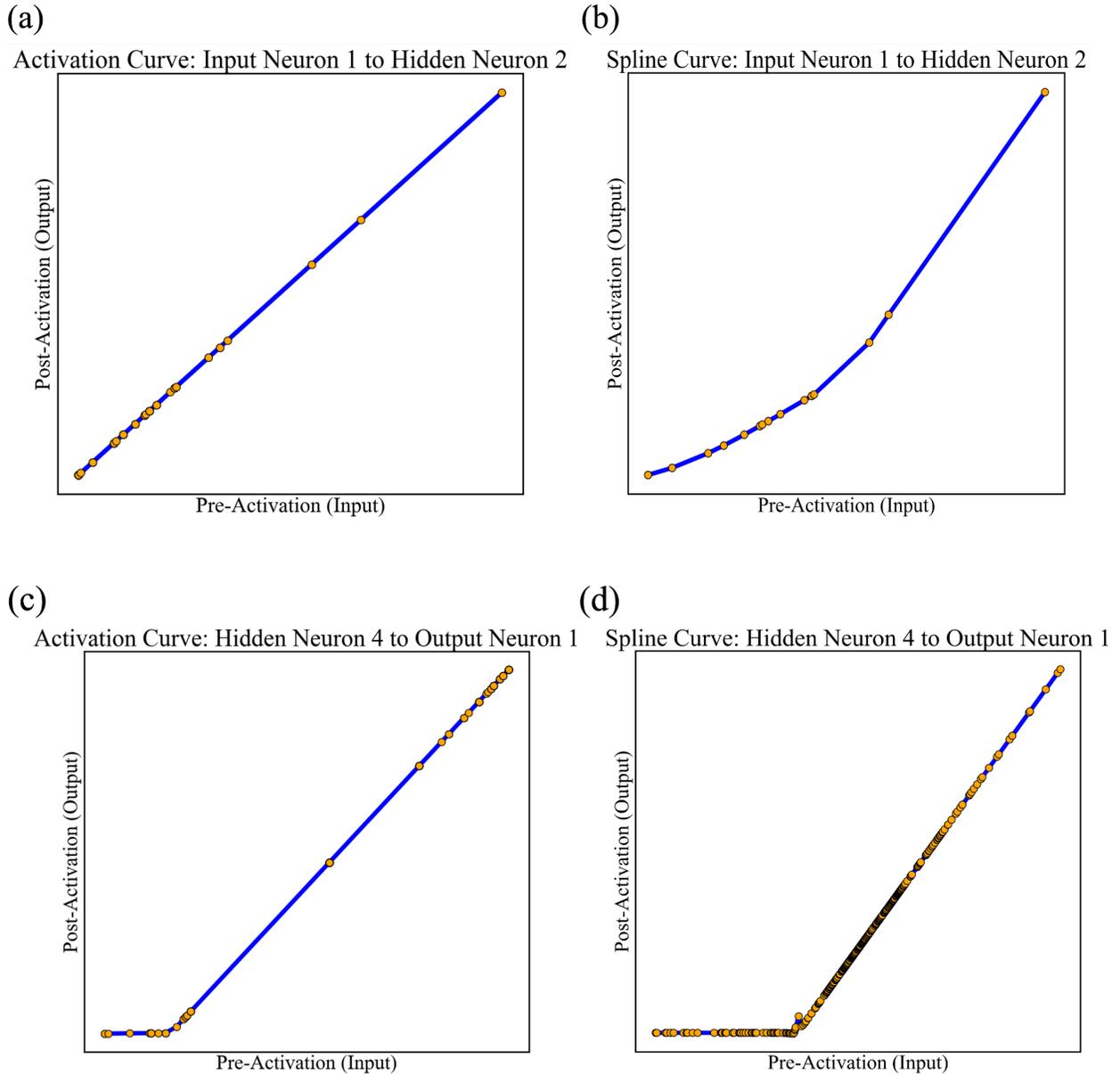

**Figure 7.** (a) MLP Activation Curve. (b) KAN spline curve of the regression model predicting ultimate tensile strength (UTS). (c) MLP Activation Curve. (d) KAN spline curve of the regression model for predicting yield strength (YS).

For the regression task, we again compare the outputs at each connection between KAN and MLP. The MLP shows a smooth, nearly linear progression between layers, using activation functions such as ReLU to introduce non-linearity. However, KAN's trainable splines show increased flexibility, capturing subtle variations in the data that the MLP overlooks. As seen from Figure 7 (a) and (b), although the trend of the MLP counterpart is similar, KAN can capture a parabolic function ($\propto x^2$) whereas the MLP can only capture the straight line ($\propto x$). Also, in certain cases like in Figure 7 (c) and (d), we can see instances where the MLP and KAN both have similar plots for the same neural connection which suggests that the learning curve for both KAN and MLP is not always very different.

**Example 3: Combined Classification and Regression**

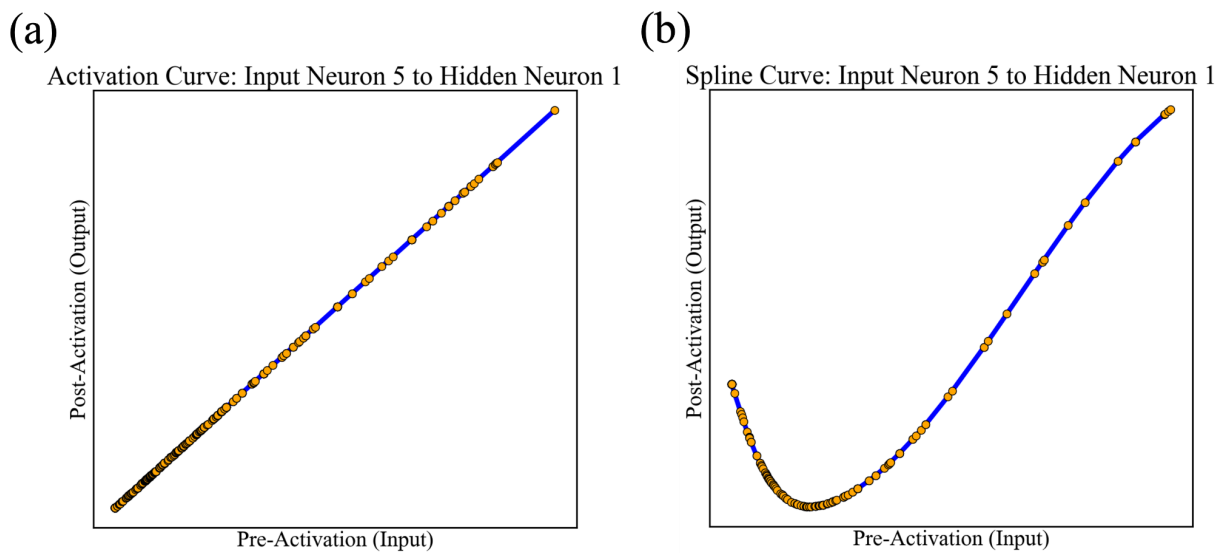

**Figure 8**: (a) MLP activation curve and (b) KAN spline curve for the regression task of predicting bulk modulus of HEAs.

In the third example, we explore a combined classification and regression task. We analyze the output functions at connections related to the regression model to predict bulk modulus. As shown in Figure 8, KAN's output demonstrates a nonlinear curve, which likely contributes to its superior performance in handling the dual task of classification and regression simultaneously. KAN's trainable non-linearity may allow it to model complex−learning curves in both classification and regression tasks better than MLP. This ability to capture complex non-linear relationships likely explains KAN's slightly superior performance across these tasks. Although this may also be the reason for overfitting of data as observed in cases like example 1 where we observe lower accuracy when applying the model to the 31 unseen HECC compounds, even though it performs well with the initial dataset.

**Conclusions**

In summary, we explore the potential of applying KAN to the design of HEAs through three examples: predicting single-phase formation, forecasting mechanical properties (yield strength and ultimate tensile strength), and classifying HEAs versus non-HEAs and predicting bulk modulus. We show that KAN exhibits stronger performance than MLPs in terms of computational accuracy. For example, KAN achieves higher classification accuracy and improves prediction accuracy for regression tasks. Our analysis reveals that KAN has an improved ability to learn complex, non-linear relationships within high-dimensional datasets, thanks to their unique combination of splines. Despite the potential of KAN, challenges remain, particularly to avoid overfitting in complex material systems exemplified by HEAs, as we observe occasional discrepancies in the classification of new HEA compounds as seen in the example of predicting single-phase formation (example 1). We expect the current work to highlight the potential of KAN to improve HEA design through more interpretable and accurate machine learning models, which will offer a possible direction for future research in the discovery of novel HEAs.

**Acknowledgements**

This work was supported by the National Science Foundation (NSF) (Grant No. DMR-2239216). This research used computational resources through the National Artificial Intelligence Research Resource (NAIRR) Pilot (NAIRR240231) and also through the Sol cluster at Arizona State University (ASU). H. A. acknowledges the SCience and ENgineering Experience (SCENE) high school program at ASU.**References**

[1] K.-H. Huang, J. Yeh, A study on the multicomponent alloy systems containing equal-mole elements, Hsinchu: National Tsing Hua University 1 (1996).
[2] M.-H. Tsai, J.-W. Yeh, High-entropy alloys: a critical review, Materials Research Letters 2(3) (2014) 107-123.
[3] E.P. George, W.A. Curtin, C.C. Tasan, High entropy alloys: A focused review of mechanical properties and deformation mechanisms, Acta Materialia 188 (2020) 435-474.
[4] E.P. George, D. Raabe, R.O. Ritchie, High-entropy alloys, Nature reviews materials 4(8) (2019) 515-534.
[5] Y. Hong, B. Hou, H. Jiang, J. Zhang, Machine learning and artificial neural network accelerated computational discoveries in materials science, Wiley Interdisciplinary Reviews: Computational Molecular Science 10(3) (2020) e1450.
[6] A.H. Naghdi, D. Massa, K. Karimi, S. Papanikolaou, High Entropy Alloy Composition Design for Mechanical Properties,  (2024).
[7] X. Wan, Z. Li, W. Yu, A. Wang, X. Ke, H. Guo, J. Su, L. Li, Q. Gui, S. Zhao, Machine learning paves the way for high entropy compounds exploration: challenges, progress, and outlook, Advanced Materials  (2023) 2305192.
[8] Y. Juan, Y. Dai, Y. Yang, J. Zhang, Accelerating materials discovery using machine learning, Journal of Materials Science & Technology 79 (2021) 178-190.
[9] D. Krishnamurthy, H. Weiland, A. Barati Farimani, E. Antono, J. Green, V. Viswanathan, Machine learning based approaches to accelerate energy materials discovery and optimization, ACS energy letters 4(1) (2018) 187-191.
[10] G. Pilania, C. Wang, X. Jiang, S. Rajasekaran, R. Ramprasad, Accelerating materials property predictions using machine learning, Scientific reports 3(1) (2013) 2810.
[11] X. Wan, W. Feng, Y. Wang, H. Wang, X. Zhang, C. Deng, N. Yang, Materials discovery and properties prediction in thermal transport via materials informatics: a mini review, Nano letters 19(6) (2019) 3387-3395.
[12] M.-C. Popescu, V.E. Balas, L. Perescu-Popescu, N. Mastorakis, Multilayer perceptron and neural networks, WSEAS Transactions on Circuits and Systems 8(7) (2009) 579-588.
[13] D.W. Ruck, S.K. Rogers, M. Kabrisky, Feature selection using a multilayer perceptron, Journal of neural network computing 2(2) (1990) 40-48.
[14] G. Vazquez, S. Chakravarty, R. Gurrola, R. Arróyave, A deep neural network regressor for phase constitution estimation in the high entropy alloy system Al-Co-Cr-Fe-Mn-Nb-Ni, npj Computational Materials 9(1) (2023) 68.
[15] Z. Zhou, Y. Zhou, Q. He, Z. Ding, F. Li, Y. Yang, Machine learning guided appraisal and exploration of phase design for high entropy alloys, npj Computational Materials 5(1) (2019) 128.27